\documentclass[tightenlines,superscriptaddress,eqsecnum,floats,nofootinbib,showpacs]
{revtex4-1}

\usepackage{amssymb}
\usepackage{stmaryrd}
\usepackage{amsmath}
\usepackage{amsfonts}
\usepackage{mathrsfs}
\usepackage{CJK}
\usepackage{amsmath,amssymb,amsfonts}
\usepackage{graphicx}

\def\be{\begin{equation}}
\def\ee{\end{equation}}
\def\ba{\begin{eqnarray}}
\def\ea{\end{eqnarray}}

\begin{document}


\title{ Fix Immirzi parameter by quasinormal modes in four and higher spacetime dimensions}

\author{Xiangdong Zhang\footnote{scxdzhang@scut.edu.cn}}
\affiliation{Department of Physics, South China University of
Technology, Guangzhou 510641, China}
\affiliation{Institute for Quantum Gravity, University of Erlangen-N{\"u}rnberg, Staudtstra{\ss}e 7 / B2, 91058 Erlangen, Germany}

\begin{abstract}
One parameter quantization ambiguity is existed in Loop quantum gravity which is called the Immirzi parameter. In this paper, we fix this free paremater by considering the quasinormal mode spectrum of black holes in four and higher spacetime dimensions. As a consequence, our result consistents with Bekenstein-Hawking entropy of a black hole. Moreover, we also give a possible quantum gravity explanation of the universal $\ln(3)$ behavior of the quasinormal mode spectrum.

\pacs{04.60.Pp, 04.70.Dy, 04.50.Gh}
\end{abstract}

\keywords{loop quantum cosmology, singularity resolution, effective equation}

\maketitle

\section{Introduction}\label{sec:introduction}

Loop quantum gravity(LQG) is a quantum gravity theory which aims to quantize General Relativity(GR) with the nonperturbative technicals\cite{Ro04,Th07,As04,Ma07}. During the past decades, many aspects of LQG have been under investigated, particularly on the subjects of loop quantum cosmology, black hole entropy and spin foam models\cite{AS11,Rovelli11,BP15}. One of the most interesting features of LQG is the existence of the discrete spectrum of the geometric operator. Such as the area and volume operator. For instance, the area spectrum is given by\cite{Ro04,Th07}
\ba
A=8\pi \gamma\ell_p^2\sqrt{j(j+1)}\label{areaspectrum4}
\ea where $j$ labels the representation of $SU(2)$ gauge group and thus takes the value of a half integer and $\gamma$ is a adjustanle real number called the Immirzi parameter. It is shown by Rovelli and Thiemann\cite{RT98}, different choices of the value of $\gamma$ lead to unitarily inequivalent quantum theory. Thus it becomes important for the LQG to fix this free parameter $\gamma$.

One way to fix Immirzi parameter $\gamma$ is to calculate the black hole entropy in the framework of LQG. In order to recover the famous Bekenstein-Hawking entropy formula \ba
S_{BH}=\frac{A}{4\ell_p^2}
\ea where $\ell_{p}=(G\hbar)^{\frac{1}{2}}$ being the Planck length, we can fix the value of $\gamma$. However, up to now, the value of $\gamma$ is yet to be uniquely fixed, different counting method gives the different value of $\gamma$\cite{BP15}. Thus it will be desired if it exists another way to fix the value of Immirzi parameter $\gamma$. Fortunately, Dreyer propose a novel way to fix the $\gamma$ by using the asymptotic behavior of the quasinormal modes(QNM) of a four dimensional Schwarzschild black hole\cite{Dreyer}. His idea is following, first he note that the highly damped real part of the quasinormal mode frequencies $\omega_{QNM}$ asymptotically approaches a fixed value \ba
\frac{\omega_{QNM}}{T}=\ln(3)
\ea where $T=\frac{1}{8\pi M}$ denotes the Hawking temperature of a four dimensional Schwarzschild black hole. This remarkable universal asymptotic behavior of four dimensional black hole first conjectured by Hod\cite{Hod98}, and confirmed analytically by Motl\cite{Motl03a}. By assuming Bohr's correspondence principle\cite{Hod98}, such an universal asymptotic frequency should be related with the one of the quanta induced by the minimum change of the quantized area of the black hole's horizon. Combining this observation with area spectrum given by Eq.(\ref{areaspectrum4}), He successfully fix the value of $\gamma$\cite{Dreyer} \ba
\gamma=\frac{\ln(3)}{2\sqrt{2}\pi}
\ea However it also forces \ba
j_{min}=1
\ea Since the loop quantum gravity in four dimensions is based on the gauge group $SU(2)$, it is thus more natural to require $j_{min}=\frac{1}{2}$ which is the minimum half integer allowed by $SU(2)$ group representation. To remove this conflict Dreyer suggests we should adapt $SO(3)$ as our gauge group rather than $SU(2)$\cite{Dreyer}. There are other explanations, for example, Corichi argued one can still keep $SU(2)$ gauge group and have consistency with QNM\cite{Corichi}. While Ling and Zhang propose that the supersymmetric extension of the LQG is preferred by QNM\cite{LZ03}.

In order to further develop this idea,  an interesting question is that, can we generalize the original Dreyer's proposal to the higher dimensions? This is of course a very difficult problem because it is well known that the analytically expression of higher dimensional QNM is very hard to obtain.  However, it is remarkablely that for the asymptotic behavior of the real part of QNM of the higher dimensional Schwarzschild black hole is already obtained analytically by Motl \cite{Motl03b}.  as \ba
\frac{\omega_{QNM}}{T}=\ln(3)
\ea where $T$ is the Hawking temperature of Schwarzschild black hole in $d+1$ dimensions. This result is also confirmed numerically in four and five dimensions\cite{CLY04}.
Such an universal asymptotic behavior of the QNM strongly hints us there might be some quantum gravity origin which yet to be discovered.

On the other hand, recently, LQG has been generalized to arbitrary spacetime dimensions by Thiemann et. al. in the series of paper\cite{BTTa,BTTb,BTTc,BTTd}. Since the LQG is based on the connection dynamics formalism of the GR. The main idea of \cite{BTTa} is that for $d+1$ dimensional GR, in order to obtain a well defined connection dynamics, one should adapt $SO(d+1)$ connections $A_a^{IJ}$ with $I,J=1,2,3\dots d$ rather than the speculated $SO(d)$ connections. With this higher dimensional connection dynamics in hand, Thiemann et. al successfully generalize the LQG to arbitrary spacetime dimensions.  Similar to the four dimensional $SU(2)$ LQG, The higher dimensional LQG also has the discrete area spectrum reads\cite{BTTc,NB13} \ba
A=8\pi\gamma\ell_p^{d-1}\sqrt{I(I+d-1)}
\ea where $I$ is an integer and $\ell_p=(G\hbar)^{\frac{1}{d-1}}$ denotes the Planck length in $d+1$ dimensions. The physical meaning of $I$ is that for every edge, we associate a simple representation of $SO(d+1)$ which is labeled by its corresponding highest weight $\Lambda=(I,0,0,...)$ with $I=1,2,3...$ being an integer. This paper thus aims to fix Immirzi parameter $\gamma$ appeared in this generalized LQG framework with spacetime dimensions $d\geq3$. Moreover, we also want to give a possible quantum gravity explanation of the universal asymptotic $\ln(3)$ behavior found in the four and higher dimensional QNM\cite{Motl03a,Motl03b,CLY04}.

This paper is organized as follows: After a short introduction, we first fix Immirzi parameter of four dimensions in Section(\ref{section1}).
Then we generalize this result to higher dimensions in Section(\ref{section2}) and explains why the universal asymptotic $\ln(3)$ behavior will be emerged due to the underlying higher dimensional LQG. Conclusions are given in the last section.

\section{Fix Immirzi parameter in four dimensions}\label{section1}

Let us first examine the four dimensional case. The four dimensional Schwarzschild solution reads \ba
ds^2=-(1-\frac{2M}{r})dt^2+(1-\frac{2M}{r})^{-1}dr^2+r^2d\Omega^2
\ea where $d\Omega^2$ is the two dimensional sphere metric and $M$ is the mass of the black hole. Using $\omega_{QNM}$ we now can fix the Immirzi parameter. Bohr's correspondence principle states that an oscillatory frequency of a classical system and a transition frequency of the corresponding quantum system should be equal\cite{Hod98}. Thus the appearance or disappearance of a puncture which labeled by the simple representation of $SO(4)$ with integer $I_{min}$ produce a transition of the quantum black hole. The area of the black hole then change as follows \ba
\Delta A=A(I_{min})=8\pi\gamma\ell_{p}^2\sqrt{I_{min}(I_{min}+2)}\label{gamma1}
\ea
Note that the change of the black hole's mass $\Delta M$ is related with its quasinormal frequency $\omega_{QNM}$ by setting \ba
\Delta M=\hbar\omega_{QNM}=\frac{\hbar\ln(3)}{8\pi M} \label{QNM}
\ea The relation of the mass $M$ and the area $A$ of a four dimensional Schwarzschild black hole reads \ba
A=16\pi M^2
\ea By using Eq.(\ref{QNM}), It is easy to see that \ba
\Delta A=32 \pi M\Delta M=4 \ln(3) \ell_p^2
\ea Comparing the above equation with Eq.(\ref{gamma1}) leads to the desired value of the Immirzi parameter \ba
\gamma=\frac{\ln(3)}{2\pi \sqrt{I_{min}(I_{min}+2)}}
\ea Now our remaining task is to fix the $I_{min}$.

Note that in four dimensions, we adapt $SO(4)$  as our gauge group, the dimension of a simple representations labeled by $\Lambda=(I_{min},0)$ with $\Lambda$ being the corresponding highest weight is $2I_{min}+1$. Thus the entropy of the horizon with $N$ punctures is given by \ba
S=N\ln(2I_{min}+1)
\ea Here we just simply assumed that the dominate contribution of the black hole entropy comes from all the punctures have the minimum value of $I_{min}$. Therefore we have \ba
N=\frac{A}{A(I_{min})}
\ea Since we already argued that $\Delta A=A(I_{min})=4 \ln(3) \ell_p^2$ and thus we have \ba
S=\frac{A\ln(2I_{min}+1)}{4 \ln(3) \ell_p^2}
\ea A complete  agreement with the Bekenstein-Hawking entropy of the black hole requires that \ba
I_{min}=1
\ea Therefore, we completely fix the Immirzi parameter \ba
\gamma=\frac{\ln(3)}{2\pi \sqrt{3}}
\ea Note that this result also consistents with group theoretical argument, since in four dimensional case, we adapt $SO(4)$ rather than $SU(2)$, thus the lowest possible choice of $I_{min}$ of the $SO(4)$ group is $I_{min}=1$. Hence the unnatural choice of $j_{min}=1$ arises in $SU(2)$ case in \cite{Dreyer} is no longer a problem in our situation.

\section{Generalization to higher dimensions}\label{section2}

In this section, we generalize the result of the last section to higher dimensions. To begin with, we first write down the $d+1$ dimensional Schwarzschild metric \ba
ds^2=-(1-\frac{m}{r^{d-2}})dt^2+(1-\frac{m}{r^{d-2}})^{-1}dr^2+r^2d\Omega^2_{d-1}
\ea where $d\Omega^2_{d-1}$ is the $d-1$ dimensional sphere metric. The ADM mass of $d+1$ dimensional Schwarzschild black hole reads \ba
M=\frac{(d-1)2\pi^{\frac{d}{2}}}{\Gamma(\frac{d}{2})}\frac{m}{16\pi G}
\ea It is easy to see that $r_+=(m)^{\frac{1}{d-2}}$ is the radius of the event horizon. The temperature and the area of the $d+1$ dimensional Schwarzschild black hole read respectively \ba
T=\frac{(d-2)\hbar}{4\pi}(m)^{-\frac{1}{d-2}},\quad \quad A_{d-1}=\frac{2\pi^{\frac{d}{2}}}{\Gamma(\frac{d}{2})}(r_+)^{d-1}=\frac{2\pi^{\frac{d}{2}}}{\Gamma(\frac{d}{2})}(m)^{\frac{d-1}{d-2}}\label{TA}
\ea The Bekenstein-Hawking entropy of the black hole is $S=\frac{A_{d-1}}{4\ell_p^{d-1}}$ with $\ell_p=(G\hbar)^{\frac{1}{d-1}}$ being the Planck length of $d+1$ dimensional spacetime. These quantity satisfy the first law of black hole thermodynamics \ba
dS=\frac{dM}{T} \label{firstlaw}
\ea

Similar with the last section, we first relate the change of the ADM mass of the black hole $\Delta M$ with quasinormal frequency $\omega_{QNM}$ as \ba
\Delta M=\hbar \omega_{QNM}=\hbar T\ln(3)\label{quasinormaln}
\ea
The area spectrum of $d+1$ dimensional spacetime reads\cite{BTTc} \ba
A=8\pi\gamma\ell_p^{d-1}\sqrt{I(I+d-1)}
\ea Thus with appearance or disappearance of a puncture with $I_{min}$, The area change of the black hole is given by \ba
\Delta A=8\pi\gamma\ell_p^{d-1}\sqrt{I_{min}(I_{min}+d-1)}\label{arean}
\ea Combining the expression of the area of the black hole (\ref{TA}) and Eq.(\ref{quasinormaln}), we find that \ba
\Delta A=4\ell_p^{d-1}\frac{\Delta M}{T}=4\ell_p^{d-1}\hbar\frac{\omega_{QNM}}{T}=4\hbar\ln(3)\ell_p^{d-1}\label{deltaA}
\ea Comparing Eq.(\ref{deltaA}) with Eq.(\ref{arean}) gives us the desired value of the Immirzi parameter \ba
\gamma=\frac{\ln(3)}{2\pi \sqrt{I_{min}(I_{min}+d-1)}}
\ea Now our remaining task is to fix the $I_{min}$.

Note that for $SO(d+1)$  group, the dimension of a simple representations $\Lambda=(I_{min},0,0,...)$ with $\Lambda$ being the corresponding highest weight is $2I_{min}+1$. Thus the entropy of the horizon with $N$ punctures is given by \ba
S=N\ln(2I_{min}+1)
\ea Here we just simply assumed that the dominate contribution comes from all the punctures have minimum value of $N_{min}$. Therefore we have \ba
N=\frac{A}{A(I_{min})}
\ea Since we already argued that $\Delta A=A(I_{min})=4 \ln(3) \ell_p^{d-1}$ and thus we have \ba
S=\frac{A\ln(2I_{min}+1)}{4 \ln(3) \ell_p^{d-1}}\label{BHentropy}
\ea Compare Eq.(\ref{BHentropy}) with the Bekenstein-Hawking entropy of the black hole requires that \ba
I_{min}=1
\ea Hence the Immirzi parameter in $d+1$ dimensions is completely fixed as follows\ba
\gamma=\frac{\ln(3)}{2\pi \sqrt{d}}
\ea Note that this result also consistents with group theoretic argument, since in $d+1$ dimensional case, we adapt $SO(d+1)$ connection dynamics rather than $SU(2)$ connection dynamics, thus the lowest admissible $I_{min}$ of the $SO(d+1)$ group is $I_{min}=1$. Hence the unnatural choice of $j_{min}=1$ appeared in $SU(2)$ case in \cite{Dreyer} does not bothers us. Moreover, since it is consistent to take $I_{min}=1$ both from the quasinormal modes and LQG perspective. Hence the dimension of the Hilbert space for one puncture with $I_{min}$ is just $d_{I_{min}}=2I_{min}+1=3$, thus the $\ln(3)$ asymptotic behavior of QNM is naturally predicted by LQG.

\section{conclusions}\label{section3}
In this paper, by relating the asymptotic behavior of the real part of QNM with area spectrum of the quantum general relativity, we successfully fix the Immirzi paraeter in $d+1$ dimensional LQG with gauge group $SO(d+1)$. Interestingly, we found that the Immirzi parameter $\gamma$ is spacetime dependent. Moreover we also provide a possible quantum gravity mechanism to explain the mysterious universal $\ln(3)$ behavior found in the asymptotic of QNM.

In the previous Dreyer's work, the quasinormal modes force $j_{min}=1$ while the LQG prefer $j_{min}=\frac{1}{2}$ because they are working on gauge group $SU(2)$. Thus people need to resort other possibility such as supersymmetry\cite{LZ03} or the contribution of $j=\frac{1}{2}$ has been suppressed\cite{Corichi} to reconcile this apparent discrepancy. However, since now we are working in the $SO(d+1)$ LQG formalism, thus it is natural to require $I_{min}=1$ from both quasinormal modes and LQG perspective. This we think can be served as one of the most atractive features of our result. Moreover, the exact value of Immirzi parameter $\gamma$ plays a crucial role when we do some phenomenological analysis of the LQG and hence becomes very useful such as in the case of higher dimensional LQC\cite{Zhang15}.

It is worth noting that there are still many issues deserve further investigation, for example, generalization to our result to supersymmetric case will be interesting. Furthermore, here we only focus on Schwarzschild black holes. We expect our scheme will be applicable to more general black holes. We hope these topics will be investigated in the near future.

\begin{acknowledgements}
The author would like to thank Norbert Bodendorfer for helpful discussions. The author would also like to thank the finical supported by CSC-DAAD postdoctoral fellowship. This work is supported by NSFC with No.11305063  and
the Fundamental Research Funds for the Central University of China.

\end{acknowledgements}


\end{document}